\documentclass{article}
\usepackage{spconf,amsmath,graphicx}

\usepackage{enumitem}
\setlist{nosep, leftmargin=14pt}

\usepackage{mwe} % to get dummy images
\usepackage{multirow} % pr tableau (pas utilisé pr le moment)
\usepackage{hyperref} % pr \autoref
\usepackage{xcolor} % pour \textcolor
\usepackage{multirow} % pour table 3 trait horizontal pas sous toutes les colonnes
\usepackage{subcaption} % à la place de subfig
\usepackage{comment}
\newcommand{\rev}[2]{\textcolor{black}{#2}}

\title{Perfusion imaging in deep prostate cancer detection from mp-MRI: can we take advantage of it?} 
\name{Audrey Duran, Gaspard Dussert and Carole Lartizien.} %\thanks{Some author footnote.}}
\address{Univ Lyon, CNRS, Inserm, INSA Lyon, UCBL, CREATIS, UMR5220, U1206, F‐69621, Villeurbanne, France}
\begin{document}
%\ninept
%
\maketitle
\begin{abstract}
%100 to 150 words

\rev{}{To our knowledge, all} deep \rev{}{computer-aided detection and diagnosis} (CAD) systems for prostate cancer (PCa) detection consider bi-parametric \rev{}{magnetic resonance imaging} (bp-MRI) only, including T2w and ADC sequences while excluding the 4D perfusion sequence,\rev{}{ which is however part of standard clinical protocols for this diagnostic task. In this paper, we question strategies to integrate information from perfusion imaging in deep neural architectures.
To do so, we} evaluate several ways to encode the perfusion information in a U-Net like architecture, also considering early versus mid fusion strategies. We compare performance of multiparametric MRI (mp-MRI) models with the baseline bp-MRI model based on a private dataset of 219 mp-MRI exams. Perfusion  maps  derived  from dynamic contrast enhanced MR exams are shown to positively impact segmentation and grading performance of PCa lesions, especially the 3D MR volume corresponding to the \textit{maximum slope} of the wash-in curve as well as \textit{Tmax}
perfusion maps. 
The latter mp-MRI models indeed outperform the bp-MRI one whatever the fusion strategy, \rev{}{with Cohen's kappa score of $0.318\pm0.019$ for the bp-MRI model and $0.378\pm0.033$ for the model including the \textit{maximum slope} with a mid fusion strategy, also achieving competitive Cohen's kappa score compared to state of the art.}

\end{abstract}
\section{Introduction}
\label{sec:intro}

Deep learning has become the state-of-the-art approach for the processing and analysis of many medical imaging problems, including detection and segmentation tasks. %\cite{tajbakhsh_embracing_2020}. 
Many CAD systems for prostate cancer (PCa) detection and segmentation from MRI are based on convolutional neural networks (CNN) \cite{wildeboer_artificial_2020}. However, deep models render difficult the inclusion of \rev{}{ high dimensional input data, such as 4-dimensional (4D) perfusion MR}.%which represent a large input.
In addition, the impact of this MR dynamic contrast enhanced (DCE) modality in PCa detection is controversial \cite{verma_overview_2012, de_visschere_dynamic_2017, woo_head--head_2018, brancato_assessment_2020}. As a consequence, as far as we know, all deep PCa segmentation models using original MR images as input (thus excluding the radiomics-based approaches) only include \rev{}{T2 weighted} (T2w) and \rev{}{diffusion weighted imaging} (DWI) sequences, leading to bi-parametric (bp) models. 
In this paper, we question the contribution of the dynamic sequence in the context of the challenging task of segmentation and grading of PCa in mp-MRI, addressed in few studies \cite{cao_joint_2019,duran2020prostate,de_vente_deep_2020,duran2022prostattention}. We evaluate several ways to encode the perfusion information in 3D and compare performance of mp-MRI models with the baseline bp-MRI model.
\section{Materials and methods}
\label{sec:matmeth}

\subsection{Perfusion maps}

\begin{figure}[h!]
\centering
  %%trim={<left> <lower> <right> <upper>} %200 250 200 250, clip, width=2.8cm,  height=2.8cm
\subfloat[Maximum slope]{\includegraphics[trim=180 260 180 260, clip, width=2.8cm, height=2.8cm]{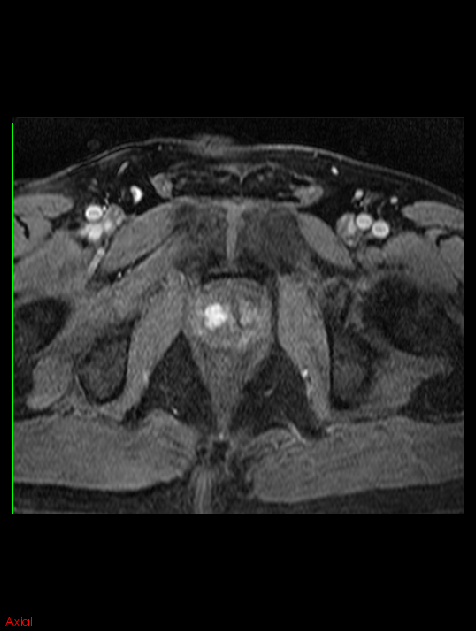}} %volume enlevé pr gagner place
\hfill
\subfloat[\% of enhancement]{\includegraphics[trim=180 260 180 260, clip, width=2.8cm,  height=2.8cm]{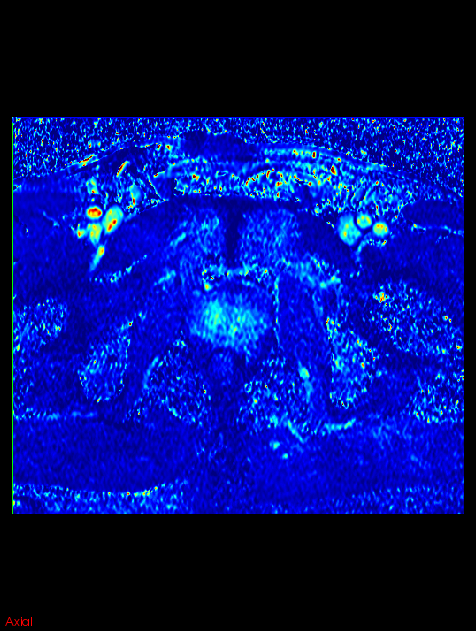}}
\hfill
\subfloat[Tmax]{\includegraphics[trim=180 260 180 260, clip, width=2.8cm, height=2.8cm]{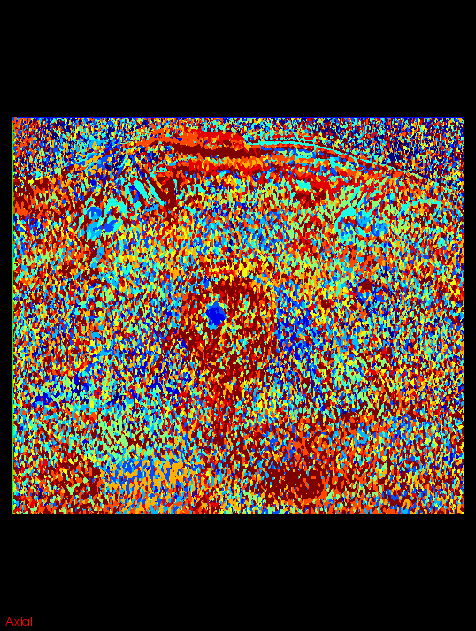}}\\
\subfloat[Wash-in slope]{\includegraphics[trim=180 260 180 260, clip, width=2.8cm, height=2.8cm]{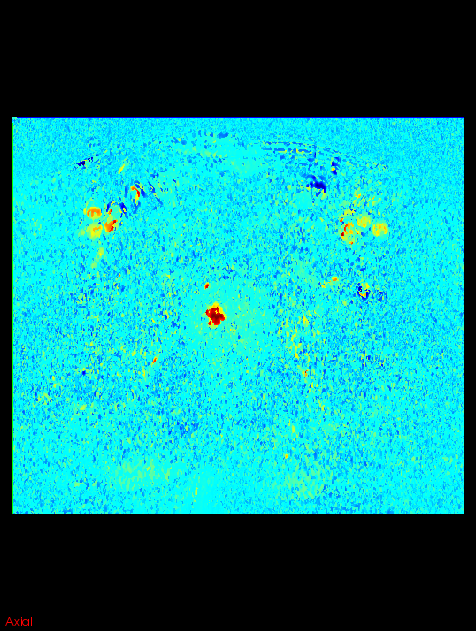}}
\hfill
\subfloat[Wash-out slope]{\includegraphics[trim=180 260 180 260, clip, width=2.8cm, height=2.8cm]{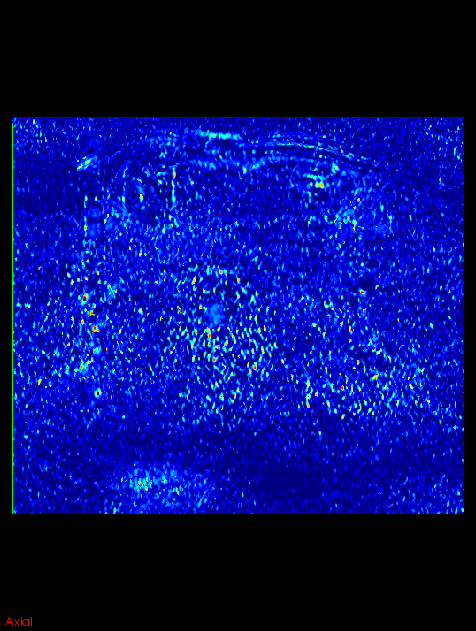}}
\hfill
\subfloat[T2w and GT]{\includegraphics[trim=180 260 180 260, clip, width=2.8cm, height=2.8cm]{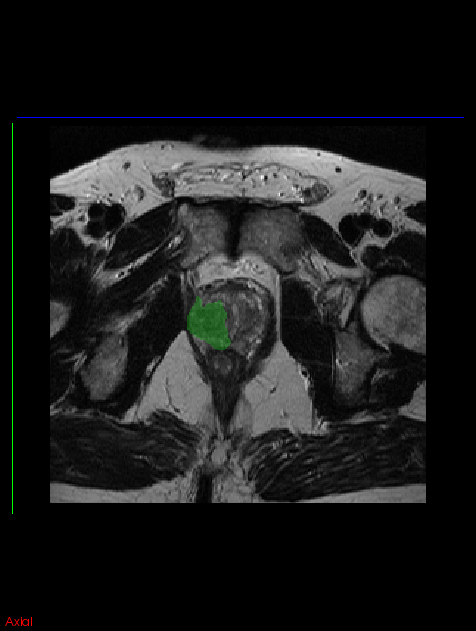}}
\caption{Example perfusion maps considered in this study. Higher values of the parametric maps ((b) to (e)) are shown in red while lower values are in blue. This example shows a GS~3+4 lesion drawn on the T2w sequence. GT: ground truth.}
\label{fig:perf}
\end{figure}

Several types of perfusion maps were derived from the time series of 3D \rev{}{dynamic contrast enhanced} (DCE) image stacks acquired for each patient after the bolus injection of gadolinium contrast agent. These maps consist either in specific 3D MR volumes extracted at specific time points from the time series or from semi-quantitative parametric maps extracted from the processing of kinetic curves at voxel level, as illustrated in \autoref{fig:perf}.\\

% \textbf{Volume at 1 minute}
% todo
\noindent \textbf{Tmax} 
\textit{Tmax} maps are parametric maps, where each voxel value corresponds to the time where the maximum value of the time intensity curve was observed.
%during the whole acquisition. 
The unit is arbitrary since dynamic MR shows variable temporal resolution. \\ % TTP = Tmax ?

%\noindent \textbf{Collaterales maps : early, mid, late} 
%todo
\noindent \textbf{Wash-in} 
The wash-in period corresponds to the period of the time intensity curve ranging from the onset to the time of peak intensity, where the onset is defined as the time corresponding to the maximum acceleration on the time-intensity curve. The \textit{wash-in} maps are derived by computing the slope observed in the time-intensity curve during the wash-in period at voxel level.
The higher the wash-in slope, the faster the wash-in speed, that is a suspicious sign of cancer tissue. \\

\noindent \textbf{Wash-out} Similarly to the wash-in map, the \textit{wash-out} map contains the voxel-based slope values observed in the time-intensity curve during the wash-out time period ranging from Tmax until the end of the acquisition.\\

\noindent \textbf{Maximum slope time volume} 
This 3D map is the 3D MR volume of the DCE times series corresponding to the volume where the maximum slope (or signal enhancement) was observed in the signal intensity curve. \\

\noindent \textbf{Maximum percentage of enhancement maps}
This 3D map reflects the percentage enhancement from the signal intensity changes between the first image and the images obtained during the wash-in period,
regardless of the time when the maximal enhancement appears. Those maps were described in \cite{yoon_dynamic_2019}. \\

Quantitative pharmacokinetic maps derived from compartmental modeling such as $K_{trans}$, $K_{ep}$, $V_e$ and $V_p$ were not considered because of their high dependence to the arterial input function and variability depending on the MR scanner acquisition parameters \cite{brunelle_variability_2018}.

\subsection{Data description}
\label{ssec:data}

The private dataset used in this study consists of a series of axial %T2 weighted (T2w), apparent diffusion coefficient (ADC) and dynamic contrast-enhanced (DCE) 
T2w, apparent diffusion coefficient (ADC) and DCE MR images from 219 patients, acquired in clinical practice at our partner clinical center.
Imaging was performed on three different scanners from different constructors and magnetic field strengths: \rev{}{67 exams on a 1.5T Symphony scanner (Siemens Medical Systems), 126 on a 3T scanner Discovery scanner (General Electric) and 26 on a 3T Ingenia scanner (Philips Healthcare)}. For DCE imaging, an intravenous injection of 0.2ml/kg of gadoterate meglumine was performed at 3 ml/s. Temporal resolution was adapted to the field strength and depends on the scanner.
All patients underwent a radical prostatectomy. After correlation with the whole-mount specimens, the uroradiologists outlined reported 338 prostate lesions as well as all prostate contours. Their distribution according to the Gleason score (GS) group is detailed in \autoref{tab:lesionsclass}, where GS 3+3 and GS $\ge$ 8  represent the less and most aggressive cancers, respectively. The detailed protocol and acquisition parameters can be found in \cite{bratan_influence_2013}. % Anonymous : \cite{bratan_anonymous}
\begin{table}[!t]
\caption{Lesions distribution by Gleason Score (GS) our dataset.} \label{tab:lesionsclass}
\centering
\begin{tabular}{ccccc}
\hline
 \textbf{ GS 3+3} & \textbf{GS 3+4} &\textbf{ GS 4+3}  & \textbf{GS $\ge$ 8} & \textbf{Total}\\ 
 104 & 126 & 56 & 52 & 338 \\
\hline
\end{tabular}
\end{table}

\subsection{Multiclass deep segmentation of PCa} 
\label{ssec:model}
\begin{figure}
    \centering
    \subfloat[Early fusion]{\includegraphics[width=\linewidth]{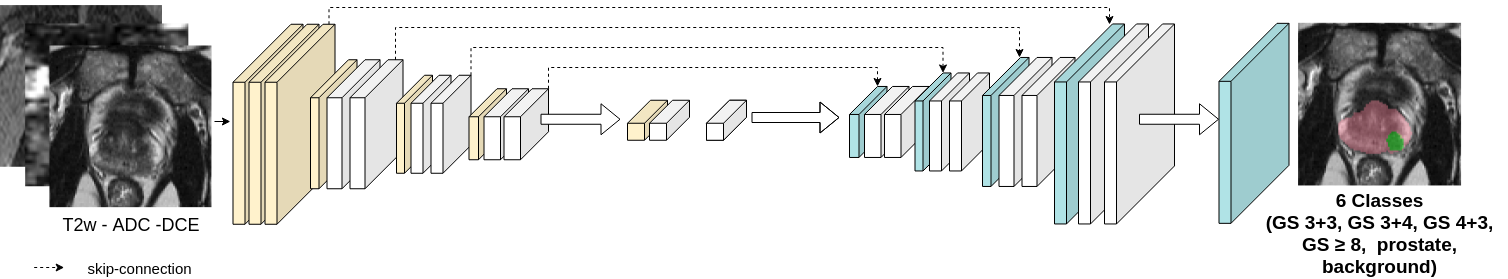}}\\
    \subfloat[Mid fusion]{\includegraphics[width=\linewidth]{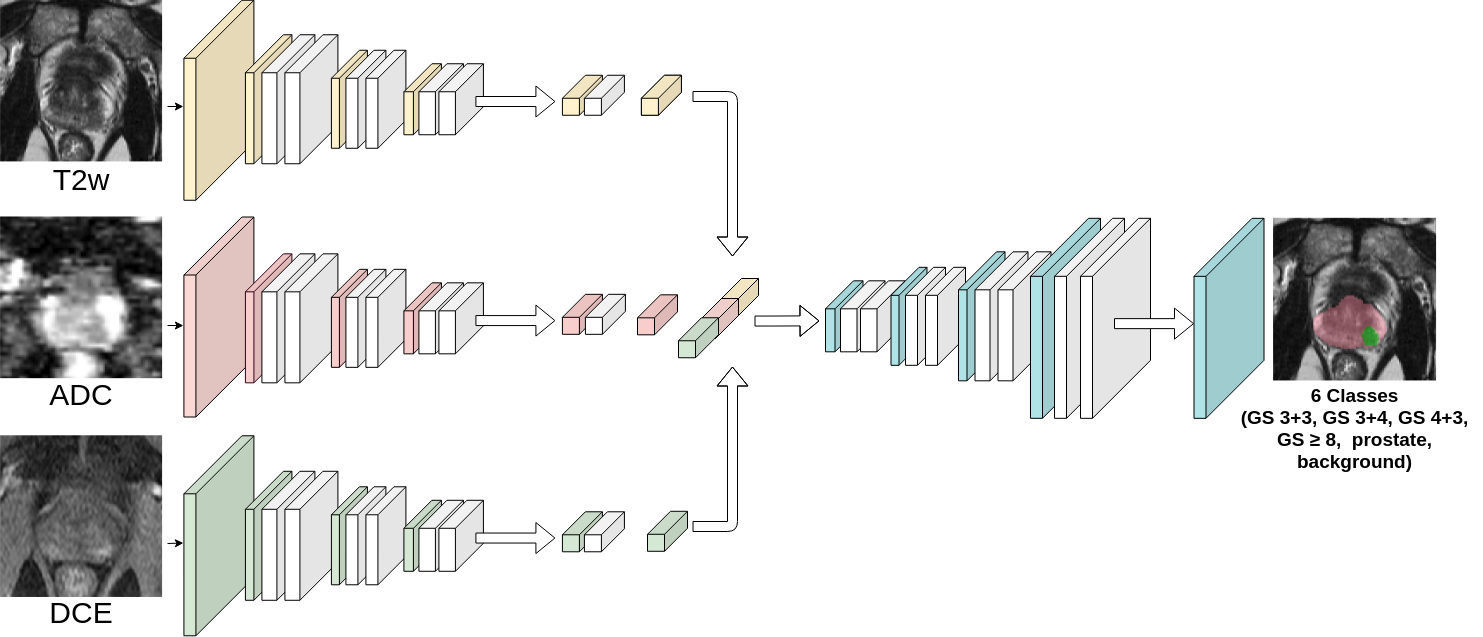}}
    \caption{\rev{}{Simplified architecture of the U-Net (a) early and (b) mid fusion models. For the sake of simplicity, skip connections are not shown for the mid fusion but the feature maps of each branch are concatenated before integration in the decoder branch.}}%Each branch skip connections are concatenating before integration in the decoder branch.}}
    \label{fig:unetfusion}
\end{figure}

The model used in this work is based on a standard four blocks U-Net \cite{ronneberger_u-net:_2015}, with batch normalization layers to reduce over-fitting and leaky ReLu activations.
It produces a 6-channels segmentation maps, corresponding to 6 class labels for the background, the overall prostate area, GS 6, GS 3+4, GS 4+3 and GS~$\ge8$ lesions. This standard U-Net architecture was shown efficient for the detection and grading of PCa with bp-MRI \cite{duran2020prostate, duran2022prostattention}.\\ % version anonymisée : duran_anonym
Two different fusion strategies of the different modalities were considered in this work as depicted on \autoref{fig:unetfusion} :
\begin{itemize}
    \item \textbf{early fusion}, where all considered modalities are stacked in a multichannel input;
    \item \textbf{mid fusion}, where each modality is encoded independently into distinct convolutional  branches of the U-Net, sharing the same decoding branch. \rev{}{Feature maps from each encoding branch are concatenated in the latent space after the encoders (after the 10th convolutional layer).} 
\end{itemize}

For each fusion strategy, we considered one baseline bi-parametric model including T2w and ADC maps as well as different multiparametric models accounting for one of the perfusion maps listed above in addition to the T2w and ADC maps.

\subsection{Experiments} 
\label{ssec:expe}
Each model was trained and validated using a 5-fold cross-validation, with 4 replicates for each cross-validation experiment for more representative results.
Multichannel input MR images were resampled to a $1\times1\times3$ mm$^3$ voxel size and automatically cropped to a $96\times96$ pixels region on the image's center. Intensity was linearly normalized into [0, 1] by volume. Data augmentation was applied during the training phase to reduce overfitting and batch size was set to 32.
All models were trained using a combination of the dice loss and cross entropy, with Adam optimizer and a L2 weight regularization with $\gamma= 10^{-4}$. The initial learning rate was set to $10^{-3}$ with a 0.5 decay after 25 epochs without validation loss improvement. The pipeline was implemented %in python 
with the Keras-Tensorflow 2.4 library.

Lesion detection performance was evaluated through free-response receiver operating characteristics (FROC) analysis, first considering the performance of each model to discriminate clinically significant (CS) lesions (GS $>$ 6), then its ability to discriminate lesions of each GS group. Cohen's quadratic weighted kappa coefficient was also computed at the lesion level as described in \cite{duran2022prostattention}. %duran_anonym}.

\section{Results}
\label{sec:results}

% * : met le tableau sur toute la longueur
\begin{table*}[h]
\resizebox{\textwidth}{!}{ 
    \centering
    \begin{tabular}{|l|c|c|c|c|c|c|} % colonnes
    \hline
    Model & Kappa & Sensi at 1FP & Sensi at 2FP & Sensi max &  Max FP & Dice prostate \\
    \hline \hline
    \multicolumn{7}{|c|}{Early fusion} \\
    \hline
        %baseline : bp-MRI & $0.306\pm0.063$& $0.526\pm0.024$& $0.666\pm0.025$& $\mathbf{0.717\pm0.022}$ & $2.866\pm0.468$ & $\mathbf{0.783\pm 0.005}$\\ (tf1)
        baseline : bp-MRI & $0.318\pm0.019$& $0.544\pm0.029$& $0.660\pm0.030$& $0.674\pm0.031$ & $2.134\pm0.139$ & $\mathbf{0.789\pm 0.002}$\\ % tf2
bp-MRI + max slope & $0.303\pm0.054$ & $\mathbf{0.598\pm0.026}$ & $0.686\pm 0.036$ & $0.705\pm0.039$ & $1.982\pm0.304$ & $0.769\pm0.007$\\
bp-MRI + \% enhancement & $0.312\pm0.035$ & $0.525\pm0.019$ & $0.649\pm0.025$ & $0.693\pm0.030$ & $2.525\pm0.185$ & $0.771\pm0.002$ \\
bp-MRI + Tmax & $0.328\pm0.026$ & $0.585\pm0.020$ & $\mathbf{0.693\pm0.024}$ & $\mathbf{0.706\pm0.019}$ & $\mathbf{1.962\pm0.165}$ & $0.770\pm0.008$ \\
bp-MRI + wash-in & $0.281\pm0.020$ & $0.541\pm0.041$ & $0.659\pm0.030$ & $0.687\pm0.024$ & $2.308\pm0.388$ & $0.784\pm0.013$ \\
bp-MRI + wash-out & $\mathbf{0.343\pm0.050}$ & $0.553\pm0.021$& $0.665\pm0.015$ & $0.680\pm0.015$ & $2.302\pm0.104$ & $0.778\pm0.003$ \\
\hline
    \multicolumn{7}{|c|}{Mid fusion} \\
    \hline
    baseline : bp-MRI & $0.333\pm0.060$ & $0.529\pm0.017$& $0.656\pm0.010$ & $0.687\pm0.010$ & $2.440\pm0.341$ &$0.792\pm0.004$\\ 
    bp-MRI + max slope & $\mathbf{0.378\pm0.033}$ & $0.569\pm0.029$ & $0.693\pm0.022$ & $0.708\pm0.017$ & $\mathbf{1.929\pm0.111}$ & $\mathbf{0.798\pm0.004}$ \\
    bp-MRI + Tmax & $0.315\pm0.064$ & $\mathbf{0.582\pm0.041}$ & $\mathbf{0.712\pm0.009}$ & $\mathbf{0.732\pm0.007}$ & $2.331\pm0.399$ & $0.778\pm0.008$\\
%\hline
%Cao et al. \cite{cao_joint_2019} & - & & $\sim 0.89$ & &  & - \\
%De Vente et al. \cite{de_vente_deep_2020} & $0.172 \pm 0.169$ & - & -& -& - & - \\
\hline
    \end{tabular}}
    \caption{Segmentation performance. Results correspond to the average metrics obtained on 4 replicates of 5-fold cross-validation. The best results for each metric and fusion strategy are in bold.} %Note that in \cite{cao_joint_2019}, only slices containing at least one lesion are included in the performance analysis so the sensitivity at a given false positive per patient rate is not comparable to ours that were estimated based on all patients slices ($\sim24$ slices per patient).}
    \label{tab:results}
\end{table*}

% TODO enlever le dice prostate ? pas l'essentiel ? 
% TODO mettre pls trucs en gras qd resultats très proche ? (ex: sensi max de max slope et Tmax)
% TODO ajouter les perfs par GS : autre tableau ou le meme ? --> mettre ce tableau dans l'autre sens pr que tout rentre ?

% Analyse: perc wash-in fait plutot baisser la variabitilité (1FP, 2FP idem, max augmente, dice prostate)
% max slope augmente plutot la variabilité, mais meilleures perfs
% sensi max et dice prostate meilleurs dans la baseline
% wash-in augment que la sens max (et fais baisser un peu le # de FP)

% * : met le tableau sur toute la longueur
\begin{table*}[h]
\resizebox{\textwidth}{!}{ 
    \centering
    \begin{tabular}{|l|cc|cc|cc|cc|}  %|c|c|c|c|c|c|c|
    \hline
      & \multicolumn{2}{c}{GS~$\ge8$} &  \multicolumn{2}{c}{GS 4+3} & \multicolumn{2}{c}{GS 3+4} & \multicolumn{2}{c|}{GS 3+3} \\
         \cline{2-9}
         Model   & 1.5 FP & 1 FP & 1.5 FP & 1 FP & 1.5 FP & 1 FP & 1.5 FP & 1 FP\\
         \hline \hline
        \multicolumn{9}{|c|}{Early fusion} \\
        \hline
        baseline : bp-MRI & $0.57\pm0.06$ & $0.56\pm0.05$ & $0.42\pm0.07$ & $0.41\pm0.06$ & $\mathbf{0.45\pm0.02}$ & $\mathbf{0.37\pm0.01}$ & $0.14\pm0.05$ & $0.12\pm0.04$\\ % tf2
        bp-MRI + max slope & $\mathbf{0.64\pm0.03}$ & $\mathbf{0.62\pm0.02}$ & $0.46\pm0.03$ & $0.45\pm0.04$ & $0.43\pm0.05$ & $0.37\pm0.02$ & $\mathbf{0.17\pm0.04}$ & $\mathbf{0.15\pm0.03}$\\
        bp-MRI + \% enhan. & $0.62\pm0.03$ & $0.61\pm0.02$ & $0.42\pm0.08$ & $0.40\pm0.06$ & $0.43\pm0.02$ & $0.36\pm0.04$& $0.06\pm0.04$ & $0.06\pm0.03$\\
        bp-MRI + Tmax & $0.61\pm0.03$ & $0.61\pm0.03$ & $\mathbf{0.47\pm0.02}$ & $\mathbf{0.46\pm0.01}$ & $0.44\pm0.03$ & $0.36\pm0.01$ & $0.09\pm0.03$ & $0.08\pm0.03$ \\
        bp-MRI + wash-in & $0.54\pm0.09$ & $0.52\pm0.05$ & $0.45\pm0.02$ & $0.45\pm0.02$ & $0.44\pm0.05$ & $0.37\pm0.05$ & $0.10\pm0.02$ & $0.09\pm0.02$\\
        bp-MRI + wash-out & $0.60\pm0.05$ & $0.59\pm0.05$ & $0.46\pm0.02$ & $0.45\pm0.03$ & $0.44\pm0.02$ & $0.37\pm0.03$ & $0.16\pm0.02$ & $0.14\pm0.02$ \\
        \hline
        \multicolumn{9}{|c|}{Mid fusion} \\
\hline
        baseline : bp-MRI & $0.56\pm0.04$ & $0.55\pm0.03$ & $0.45\pm0.10$ & $0.44\pm0.09$ & $0.46\pm0.02$ & $0.38\pm0.04$ & $\mathbf{0.20\pm0.02}$ & $\mathbf{0.17\pm0.02}$\\
        bp-MRI + max slope & $\mathbf{0.58\pm0.05}$ & $\mathbf{0.56\pm0.03}$ & $0.37\pm0.03$ & $0.37\pm0.03$ & $0.47\pm0.02$ & $0.39\pm0.02$ & $0.17\pm0.03$ & $0.16\pm0.04$ \\
        bp-MRI + Tmax & $\mathbf{0.58\pm0.05}$ & $0.54\pm0.03$ & $\mathbf{0.51\pm0.03}$ & $\mathbf{0.49\pm0.04}$ & $\mathbf{0.50\pm0.03}$ & $\mathbf{0.43\pm0.06}$ & $0.11\pm0.02$ & $0.10\pm0.02$\\
 \hline
    \end{tabular}}
    \caption{Mean detection sensitivity for each GS group. Results correspond to the average metrics obtained on 4 replicates of 5-fold cross-validation. The best results for each metric and fusion strategy are in bold.}
    \label{tab:results_gs}
\end{table*}

% TODO mettre moins de précision pr gagner de la place ? 10-2 plutot
% TODO ne garder que à un des 2 # de FP ? tjrs le meme qui a les meilleures perfs
\begin{figure*}[h!]
  %trim={<left> <lower> <right> <upper>}
  \centering
\subfloat[ground truth]{\includegraphics[trim=30 100 30 150, clip, width=0.18\linewidth]{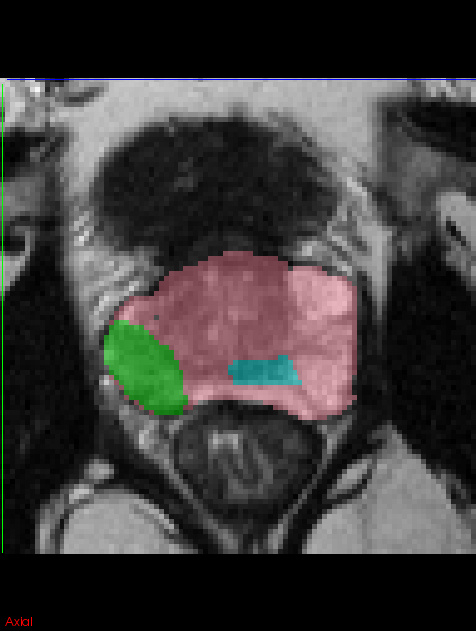}}
\hfill
\subfloat[baseline : bp-MRI]{\includegraphics[trim=30 100 30 150, clip,width=0.18\linewidth]{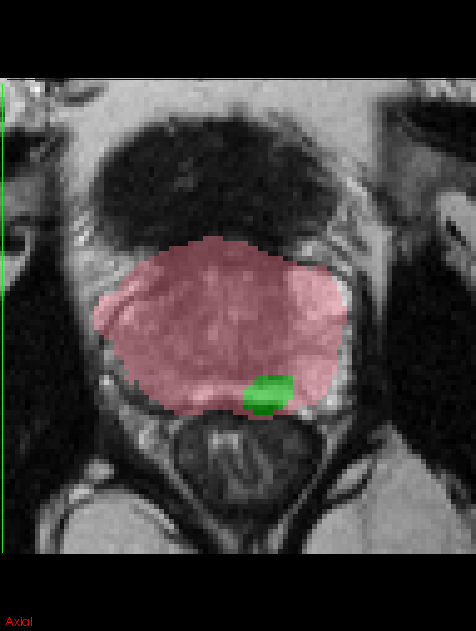}}
\hfill
\subfloat[bp-MRI + max slope]{\includegraphics[trim=30 100 30 150, clip,width=0.18\linewidth]{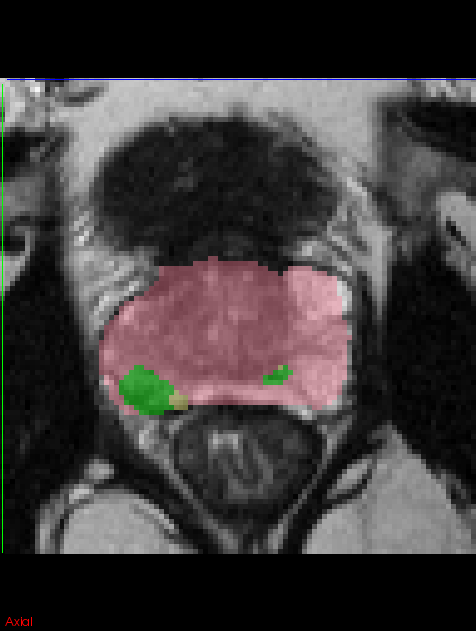}}
\hfill
\subfloat[bp-MRI + Tmax]{\includegraphics[trim=30 100 30 150, clip,width=0.18\linewidth]{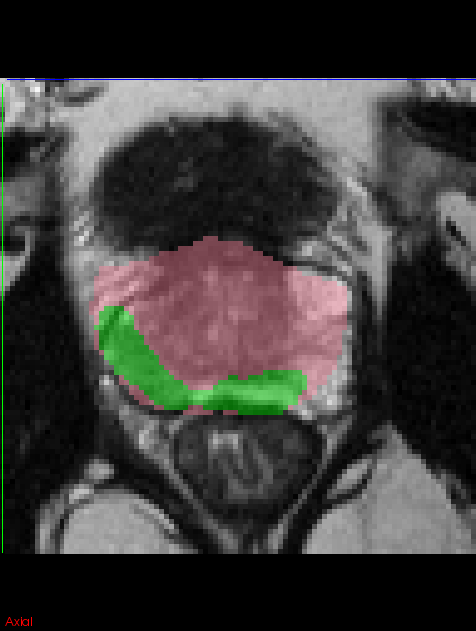}}
\hfill
\subfloat[bp-MRI + wash-out]{\includegraphics[trim=30 100 30 150, clip,width=0.18\linewidth]{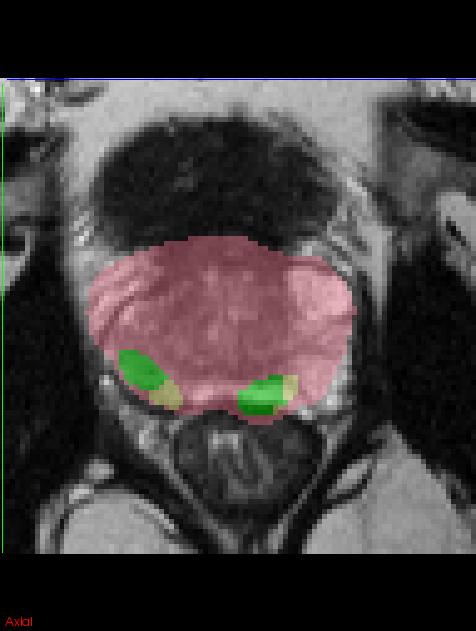}} % 0.19
\caption{Example predictions for the same slice for several models. The two outlined regions in the ground truth image correspond to GS~3+4 (green) and GS~4+3 (blue) PCa lesions, respectively.}
\label{fig:pred}
\end{figure*}

\autoref{tab:results} shows detection and segmentation performance of CS (GS $>$ 6) lesions for each perfusion map and each fusion strategy in comparison with the baseline bp-MRI model. Considering the early fusion strategy, the addition of the \textit{max slope} volume, \textit{Tmax} or \textit{wash-out} map to the T2 and ADC maps is shown to boost the model performance compared to the baseline. For the \textit{max slope} and the \textit{Tmax} maps, we observe a sensitivity gain of 5.4\% and 4.1\% respectively at 1 false positive (FP), 2.6\% and 3.3\% at 2FP and the maximum sensitivity increased by 5.8\% and 5.9\%. For those 2 models, the maximal number of FP is also reduced by 0.152 for the \textit{max slope} and 0.172 for the \textit{Tmax}. \textit{\% enhancement} and \textit{wash-in} maps are not shown to bring discriminant information for this task.
Considering the Cohen's kappa coefficient, the best value of $0.343\pm0.050$ is obtained with the \textit{wash-out} model, that outperforms the baseline ($0.318\pm0.019$). 

Performance achieved with the two best performing parametric maps (\textit{Tmax} and \textit{max slope} volume) of the early fusion scenario were compared to that achieved with the same perfusion maps in the mid-fusion strategy. This approach does not show a clear impact : at 1FP, sensitivities of these 3 models (second part of \autoref{tab:results}) are lower than with the early fusion strategy, but at 2FP, the \textit{max slope} and \textit{Tmax} sensitivies are higher than with the early fusion strategy. 

\autoref{tab:results_gs} allows a finer analysis of the sensitivity for each Gleason Score (GS) group. It reflects the model ability to both localize lesions and assign their correct GS. 
Regarding the early fusion strategy, here again, models trained with \textit{Tmax} perfusion map or \textit{max slope} volume 
outperform the baseline, except for the GS~3+4 (class with the highest number of lesions in the dataset, see \autoref{tab:lesionsclass}) where sensitivities of all models are very close and for the GS~3+3 for \textit{Tmax}. The \textit{wash-out} is also shown to perform well and outperform the baseline for each GS but the GS~3+4 at 1.5FP. 
\rev{}{The mid fusion strategy seems to outperform the early fusion, in particular when accounting for \textit{Tmax} perfusion map.}

%\vfill
%\pagebreak
\section{Conclusions}
\label{sec:conclu}

This study demonstrates that perfusion maps derived from DCE MR exams positively impact the performance of deep multiclass segmentation models of PCa. Reported performance are in par with the state of the art and outperform the reported kappa of $0.172\pm0.169$ \cite{de_vente_deep_2020} with a bp-MRI ordinal regression model \rev{}{on the ProstateX-2 challenge dataset}. 
Performance gain was shown to depend on the considered perfusion maps and fusion strategy. Although no strict guidelines could be driven from this study, we showed that, globally, the \textit{Tmax},
and \textit{max slope} volume improved detection sensitivity of CS and GS PCa lesions. This is concordant with other studies where the \textit{Tmax} was found to have a significant impact on the detection of PCa lesions \cite{zhao_dynamic_2021}. The mid fusion strategy seems to be beneficial both in terms of detection sensitivity and prostate segmentation. \rev{}{The concatenation of each sequence data at a higher representation level is likely to allow extracting more discriminant features from each MR sequence. This fusion strategy might also be less sensitive to small misalignment between sequences.}

Our study does not demonstrate the positive impact of \textit{wash-in} maps unlike some other standard radiomic studies \cite{sung_prostate_2011, hoang_dinh_quantitative_2016}. This might be explained by the way we extracted \textit{wash-in} maps : the end of the wash-in period was indeed defined at the time where the brightest intensity was observed in the voxel, that might be reached at the end of the acquisition.
In addition, the intensity normalisation was performed after the volume slices were cropped to a $96\times96$ size, so it doesn't consider the external iliac vein and artery, where a high signal is observed at the arrival time. This might induce high inter-variability of the \textit{wash-in} maps impairing the impact of this perfusion map.

Perspectives would be to further evaluate the impact of the perfusion map normalisation as well as include together the three perfusion maps (\textit{Tmax}, \textit{max slope}, \textit{wash-out}) in addition to the T2w and ADC maps in the U-Net based PCa segmentation model, in a early or mid fusion strategy.


% Below is an example of how to insert images. Delete the ``\vspace'' line,
% uncomment the preceding line ``\centerline...'' and replace ``imageX.ps''
% with a suitable PostScript file name.
% -------------------------------------------------------------------------

%\vfill
%\pagebreak
%\begin{samepage}
\section{Compliance with ethical standards}
\label{sec:ethics}
This study was performed in line with the principles of the Declaration of Helsinki. 
% Pas anonymisé
Approval was granted by the appropriate national administrative authorities (\textit{Comité de Protection des Personnes, reference L 09-04} and \textit{Commission Nationale de l’Informatique et des Libertés, treatment n$^{\circ}$ 08-06}) and patients gave written informed consent for researchers to use their MR imaging and pathologic data.
%\vfill\pagebreak
%\nopagebreak
% Exemples
%\vspace{-2em}
%\input{section/6_Acknowledgments}
\section{Acknowledgments}
%\vspace{-1em}
\label{sec:acknowledgments}
% Pas anonymisé
This work was supported by the RHU PERFUSE (ANR-17-RHUS-0006) of Université Claude Bernard Lyon 1 (UCBL), within the program “Investissements d'Avenir” operated by the French National Research Agency (ANR).
%\vfill
%\end{samepage}

% \vfill
% \pagebreak
\bibliographystyle{IEEEbib}
\bibliography{strings,refs}

\begin{thebibliography}{10}

\bibitem{wildeboer_artificial_2020}
Rogier~R. Wildeboer, Ruud J.~G. van Sloun, Hessel Wijkstra, and Massimo Mischi,
\newblock ``Artificial intelligence in multiparametric prostate cancer imaging
  with focus on deep-learning methods,''
\newblock {\em Computer Methods and Programs in Biomedicine}, vol. 189, pp.
  105316, 2020.

\bibitem{verma_overview_2012}
Sadhna Verma, Baris Turkbey, Naira Muradyan, Arumugam Rajesh, Francois Cornud,
  Masoom~A. Haider, Peter~L. Choyke, and Mukesh Harisinghani,
\newblock ``Overview of {Dynamic} {Contrast}-{Enhanced} {MRI} in {Prostate}
  {Cancer} {Diagnosis} and {Management},''
\newblock {\em American Journal of Roentgenology}, vol. 198, no. 6, pp.
  1277--1288, June 2012,
\newblock Publisher: American Roentgen Ray Society.

\bibitem{de_visschere_dynamic_2017}
P.~De~Visschere, N.~Lumen, P.~Ost, K.~Decaestecker, E.~Pattyn, and G.~Villeirs,
\newblock ``Dynamic contrast-enhanced imaging has limited added value over
  {T2}-weighted imaging and diffusion-weighted imaging when using {PI}-{RADSv2}
  for diagnosis of clinically significant prostate cancer in patients with
  elevated {PSA},''
\newblock {\em Clinical Radiology}, vol. 72, no. 1, pp. 23--32, Jan. 2017.

\bibitem{woo_head--head_2018}
Sungmin Woo, Chong~Hyun Suh, Sang~Youn Kim, Jeong~Yeon Cho, Seung~Hyup Kim, and
  Min~Hoan Moon,
\newblock ``Head-to-{Head} {Comparison} {Between} {Biparametric} and
  {Multiparametric} {MRI} for the {Diagnosis} of {Prostate} {Cancer}: {A}
  {Systematic} {Review} and {Meta}-{Analysis},''
\newblock {\em American Journal of Roentgenology}, vol. 211, no. 5, pp.
  W226--W241, Nov. 2018,
\newblock Publisher: American Roentgen Ray Society.

\bibitem{brancato_assessment_2020}
Valentina Brancato, Giuseppe Di~Costanzo, Luca Basso, Liberatore Tramontano,
  Marta Puglia, Alfonso Ragozzino, and Carlo Cavaliere,
\newblock ``Assessment of {DCE} {Utility} for {PCa} {Diagnosis} {Using}
  {PI}-{RADS} v2.1: {Effects} on {Diagnostic} {Accuracy} and
  {Reproducibility},''
\newblock {\em Diagnostics}, vol. 10, no. 3, pp. 164, Mar. 2020,
\newblock Number: 3 Publisher: Multidisciplinary Digital Publishing Institute.

\bibitem{cao_joint_2019}
Ruiming Cao, Amirhossein Mohammadian~Bajgiran, Sohrab Afshari~Mirak, Sepideh
  Shakeri, Xinran Zhong, Dieter Enzmann, Steven Raman, and Kyunghyun Sung,
\newblock ``Joint {Prostate} {Cancer} {Detection} and {Gleason} {Score}
  {Prediction} in mp-{MRI} via {FocalNet},''
\newblock {\em IEEE Transactions on Medical Imaging}, 2019.

\bibitem{duran2020prostate}
Audrey Duran, Pierre-Marc Jodoin, and Carole Lartizien,
\newblock ``Prostate {Cancer} {Semantic} {Segmentation} by {Gleason} {Score}
  {Group} in bi-parametric {MRI} with {Self} {Attention} {Model} on the
  {Peripheral} {Zone},''
\newblock in {\em Medical {Imaging} with {Deep} {Learning}}. Sept. 2020, pp.
  193--204, PMLR.

\bibitem{de_vente_deep_2020}
Coen De~Vente, Pieter Vos, Matin Hosseinzadeh, Josien Pluim, and Mitko Veta,
\newblock ``Deep {Learning} {Regression} for {Prostate} {Cancer} {Detection}
  and {Grading} in {Bi}-parametric {MRI},''
\newblock {\em IEEE Trans. Biomed. Eng.}, pp. 1--1, 2020.

\bibitem{duran2022prostattention}
Audrey Duran, Gaspard Dussert, Olivier Rouvière, Tristan Jaouen, Pierre-Marc
  Jodoin, and Carole Lartizien,
\newblock ``{ProstAttention-Net}: a deep attention model for prostate cancer
  segmentation by aggressiveness in {MRI} scans,''
\newblock {\em Medical Image Analysis}, p. 102347, 2022.

\bibitem{yoon_dynamic_2019}
Ji~Yoon, Moon Choi, Young-Joon Lee, and Seung Jung,
\newblock ``Dynamic {Contrast}-{Enhanced} {MRI} of the {Prostate}: {Can}
  {Auto}-{Generated} {Wash}-in {Color} {Map} {Be} {Useful} in {Detecting}
  {Focal} {Lesion} {Enhancement}?,''
\newblock {\em Investigative Magnetic Resonance Imaging}, vol. 23, pp. 220,
  Jan. 2019.

\bibitem{brunelle_variability_2018}
S.~Brunelle, C.~Zemmour, F.~Bratan, F.~Mège-Lechevallier, A.~Ruffion,
  M.~Colombel, S.~Crouzet, A.~Sarran, and O.~Rouvière,
\newblock ``Variability induced by the {MR} imager in dynamic contrast-enhanced
  imaging of the prostate,''
\newblock {\em Diagnostic and Interventional Imaging}, vol. 99, no. 4, pp.
  255--264, Apr. 2018.

\bibitem{bratan_influence_2013}
Flavie Bratan, Emilie Niaf, Christelle Melodelima, Anne~Laure Chesnais, Rémi
  Souchon, Florence Mège-Lechevallier, Marc Colombel, and Olivier Rouvière,
\newblock ``Influence of imaging and histological factors on prostate cancer
  detection and localisation on multiparametric {MRI}: a prospective study,''
\newblock {\em Eur Radiol}, vol. 23, no. 7, pp. 2019--2029, July 2013.

\bibitem{ronneberger_u-net:_2015}
Olaf Ronneberger, Philipp Fischer, and Thomas Brox,
\newblock ``U-{Net}: {Convolutional} {Networks} for {Biomedical} {Image}
  {Segmentation},''
\newblock {\em arXiv:1505.04597 [cs]}, 2015.

\bibitem{zhao_dynamic_2021}
Jing Zhao, Avan Kader, Dilyana~B. Mangarova, Julia Brangsch, Winfried Brenner,
  Bernd Hamm, and Marcus~R. Makowski,
\newblock ``Dynamic {Contrast}-{Enhanced} {MRI} of {Prostate} {Lesions} of
  {Simultaneous} [{68Ga}]{Ga}-{PSMA}-11 {PET}/{MRI}: {Comparison} between
  {Intraprostatic} {Lesions} and {Correlation} between {Perfusion}
  {Parameters},''
\newblock {\em Cancers}, vol. 13, no. 6, pp. 1404, Jan. 2021,
\newblock Number: 6 Publisher: Multidisciplinary Digital Publishing Institute.

\bibitem{sung_prostate_2011}
Yu~Sub Sung, Heon-Ju Kwon, Bum-Woo Park, Gyunggoo Cho, Chang~Kyung Lee,
  Kyoung-Sik Cho, and Jeong~Kon Kim,
\newblock ``Prostate {Cancer} {Detection} on {Dynamic} {Contrast}-{Enhanced}
  {MRI}: {Computer}-{Aided} {Diagnosis} {Versus} {Single} {Perfusion}
  {Parameter} {Maps},''
\newblock {\em American Journal of Roentgenology}, vol. 197, no. 5, pp.
  1122--1129, Nov. 2011,
\newblock Publisher: American Roentgen Ray Society.

\bibitem{hoang_dinh_quantitative_2016}
Au~Hoang~Dinh, Christelle Melodelima, R{\'e}mi Souchon, J{\'e}r{\^o}me Lehaire,
  Flavie Bratan, Florence M{\`e}ge-Lechevallier, Alain Ruffion, S{\'e}bastien
  Crouzet, Marc Colombel, and Olivier Rouvi{\`e}re,
\newblock ``Quantitative {Analysis} of {Prostate} {Multiparametric} {MR}
  {Images} for {Detection} of {Aggressive} {Prostate} {Cancer} in the
  {Peripheral} {Zone}: {A} {Multiple} {Imager} {Study},''
\newblock {\em Radiology}, vol. 280, no. 1, pp. 117--127, Feb. 2016.

\end{thebibliography}

\end{document}